\documentclass[aps,prd,twocolumn,preprintnumbers,superscriptaddress,floatfix]{revtex4}

\usepackage{multirow, graphicx,amssymb,url,mathrsfs,amsmath}
\usepackage{eucal,wrapfig,boxedminipage,setspace,subfigure}
\usepackage{amsxtra,amstext,latexsym,dsfont}

\def\IR{{\hbox{{\rm I}\kern-.2em\hbox{\rm R}}}}
\def\IB{{\hbox{{\rm I}\kern-.2em\hbox{\rm B}}}}
\def\IN{{\hbox{{\rm I}\kern-.2em\hbox{\rm N}}}}
\def\IC{\,\,{\hbox{{\rm I}\kern-.59em\hbox{\bf C}}}}
\def\IZ{{\hbox{{\rm Z}\kern-.4em\hbox{\rm Z}}}}
\def\IP{{\hbox{{\rm I}\kern-.2em\hbox{\rm P}}}}
\def\IH{{\hbox{{\rm I}\kern-.4em\hbox{\rm H}}}}
\def\ID{{\hbox{{\rm I}\kern-.2em\hbox{\rm D}}}}

\newcommand{\beq}{\begin{equation}}
\newcommand{\eeq}{\end{equation}}
\newcommand{\bea}{\begin{eqnarray}}
\newcommand{\eea}{\end{eqnarray}}

\begin{document}

\voffset 1cm

\newcommand\sect[1]{\emph{#1}---}

\title{Soft Walls in Dynamic AdS/QCD and the Techni-dilaton}

\author{Nick Evans}
\affiliation{STAG Research Centre \&  Physics and Astronomy, University of
Southampton, Southampton, SO17 1BJ, UK}

\author{Peter Jones}
\affiliation{STAG Research Centre \&  Physics and Astronomy, University of
Southampton, Southampton, SO17 1BJ, UK}

\author{Marc Scott}
\affiliation{STAG Research Centre \&  Physics and Astronomy, University of
Southampton, Southampton, SO17 1BJ, UK}

\begin{abstract}
Dynamic AdS/QCD is a modification of AdS/QCD that includes the running of the anomalous dimension of the $\bar{q} q$ quark bilinear and in which the generation of the constituent quark mass plays the role of an IR wall. The model allows one to move away smoothly from the controlled spectrum of the ${\cal N}=2$ super Yang-Mills theory of the D3/probe-D7 system to more QCD-like theories with chiral symmetry breaking. We investigate soft wall behaviour in the model that gives Regge trajectories with $M_{n,s}^2 \sim n, s$. To achieve these behaviours requires the quark's constituent mass to fall peculiarly sharply in the IR so that meson physics is sensitive to RG scales well below the quark's on-shell mass. Including soft wall behaviour in models of walking gauge dynamics breaks the near conformal symmetry which is present above the quark on-shell mass which can generate a large mass for the techni-dilaton like state. We conclude that the meson spectrum is rather sensitive to the IR  decoupling. 
\noindent

\end{abstract}

\maketitle

\newpage

\section{Introduction}

The AdS/CFT Correspondence \cite{Maldacena:1997re} combined with the introduction of flavour branes \cite{Karch:2002sh,review} provides a quantitative method for computation of the physics of quarks in at least some strongly coupled gauge theories. For example, the light meson spectrum of the ${\cal N}=2$ theory with a small number of quark multiplets in the background of ${\cal N}=4$ super-Yang-Mills is known \cite{Kruczenski}. It has been natural to attempt to move these techniques towards more QCD-like theories \cite{Babington:2003vm,Kruczenski:2003uq,Sakai:2004cn,Filev:2007gb} although a more phenomenological bottom-up methodology has been the only way to directly proceed to QCD. AdS/QCD \cite{Erlich:2005qh,DaRold:2005zs} is the broadest brush-stroke example, being the maximally simplified model in the spirit of a probe D7 action embedded in AdS$_5$. It was recognised early on though that the model suffers from radially excited states whose masses grow as $n$ (the excitation number) \cite{Schreiber:2004ie,Shifman:2005zn} and, if minimally extended to include higher spin states, masses that grow as $s$ (the spin). In QCD, meson Regge trajectories typically show $\sqrt{n}$ and $\sqrt{s}$ behaviour and indeed this was the original motivation for string theory. In AdS/CFT the ultra strong coupling limit leaves a classical supergravity theory for the lightest states without strings in the AdS space to match the expectation that in QCD quarks are joined by extended gluonic flux tubes. 

In \cite{Karch:2006pv} the authors pointed out that modifications of the infra-red (IR) AdS geometry or the IR behaviour of a dilaton field in AdS could be used to achieve the expected Regge form for the masses. This clever trick undoes the argument that a dual field theory can't describe this aspect of the QCD spectrum. However, the trick is not completely convincing since the mesonic spectrum is determined by modifications of the theory in the deep IR well below the constituent quark mass and confinement scale which appears a peculiar violation of decoupling. Nevertheless such IR modifications have been incorporated into models such as \cite{Gursoy:2007cb,Jarvinen:2011qe} that provide a remarkably good description of QCD (see also for example the recent \cite{Ballon-Bayona:2015wra}).

Here we will present an analysis in which we introduce soft-wall behaviour in the Dynamic AdS/QCD model \cite{Alho:2013dka}. The model is a slightly more sophisticated version of AdS/QCD retaining a few more features of probe brane embeddings in rigorous string theory settings. In particular it allows one to smoothly move (in a phenomenological rather than rigorous way) from the ${\cal N}=2$ theory to more QCD-like behaviour. It also provides an interpretation of the chiral symmetry breaking condensation as the dynamical generation of a IR quark mass.  We will see that to introduce a soft wall and obtain a meson spectrum requires a rather peculiar profile for the dynamical mass (certainly not seen in top down models) in order to allow physics in the deep IR to enter the meson physics - in particular the dynamical quark mass must vanish in the deep IR. This analysis provides some support for those who claim there is an inherent tension in the use of non-stringy descriptions of QCD states. 

Recent work \cite{Jarvinen:2011qe, Kutasov:2012uq, Alho:2013dka} has considered extending AdS/QCD models beyond just QCD to theories with arbitrary $N_f$ and $N_c$ including theories speculated to run to IR fixed points and to behave as walking theories \cite{Holdom:1981rm}. These models hopefully provide guidance to lattice practitioners who are attempting to simulate such theories \cite{Aoki:2012ep,Cheng:2013eu,Deuzeman:2011pa,Appelquist:2011dp,Hasenfratz:2010fi,Fodor:2009wk,Aoki:2012eq,Appelquist:2009ty,Deuzeman:2009mh,Shamir:2008pb,Iwasaki:2003de} and one hopes that broad trends in the behaviours of meson masses as one enters this regime will be correctly displayed. The key extra ingredient beyond the simplest AdS/QCD models is to allow the AdS mass squared of the scalar which describes the quark condensate (and encodes its dimension via $M^2 = \Delta(\Delta-4)$), to run with RG scale. Chiral symmetry breaking occurs at the scale where the Breitenlohner Freedman (BF) bound \cite{Breitenlohner:1982jf} is violated ($M^2=-4$ in AdS$_5$ corresponding to $\Delta =2$, or an anomalous dimension for $\bar{q} q$ of $\gamma=1$). 

An interesting debate about walking theories is whether the pseudo-conformal regime generates a bound state in the spectrum that is anomalously light since it is a Goldstone for conformal symmetry breaking. Interestingly in Dynamic AdS/QCD the spin zero $\bar{q}q$ $\sigma$ or $f_0$ meson was observed to become light \cite{Alho:2013dka,Evans:2013vca} (the models of \cite{Lawrance:2012cg} also describe a light dilaton holographically). On the other hand in the model of \cite{Jarvinen:2011qe} its mass did not fall relative to other states in this regime.  Here we wish to suggest that the key difference between these models is that the latter incorporates a soft wall IR (the IR behaviour of their tachyon field is crucial to this dynamics also) whilst the former don't. To test this we show that if soft wall behaviour (however artificial) is introduced into the Dynamic AdS/QCD model then the shift to the IR behaviour of the quark mass introduces rather strong conformal symmetry breaking and the resulting $\sigma$ becomes heavy. This at least makes it clear that the behaviour of this state is sensitive to how the mesonic physics decouples at strong coupling at scales beneath the scale of the dynamically generated quark mass.

\section{hard \& Soft wall AdS/QCD}

We begin by reviewing briefly the hard \cite{Erlich:2005qh,DaRold:2005zs} and soft wall \cite{Karch:2006pv} AdS/QCD models. 
We assume the $s=1$ $\rho$ meson is created by the operator $\bar{q} \gamma^\mu q$. Higher Regge states of greater spin are associated with the operators $\bar{q} \gamma^\mu \partial^\nu q$,
$\bar{q} \gamma^\mu \partial^\nu \partial^\lambda q$,.. and so on with $s-1$ derivatives inserted. The operators have dimension $2+s$.

Holographically each of these operators will be associated with an $s$ index  field $A^{\mu ...}$ in the dual geometry. 
In the hard wall model the dilaton $\Phi$ is a constant and one uses the AdS$_5$ metric
\begin{equation} 
ds^2 = r^2 dx_4^2 + {1 \over r^2} dr^2 \end{equation}
$r$ has the usual interpretation as an energy scale and is subject to a sharp cut-off at some $r_0$.

The action is then
\begin{equation} S \sim \int d^4x dr ~~ \sqrt{-g} e^{-\Phi} {r^{(4s-4)} \over 4 g_5^2} (\partial^\mu A^{\nu \lambda...})^2 \end{equation}
$\Phi$ is a dilaton field. Note the factors of $r$ are present since we have rescaled the fields $A^{\nu ...}$ to have dimension $2-s$. Writing such an action implicitly assumes a prescription for higher spin fields in AdS - here we have written the most naive possibility.

To find the bound state masses we consider, for example, states polarized in $x,y$ and moving in $z$. We seek linearized solutions of the form $V(r) \epsilon^{\mu...} e^{-ik.x}$ with $k^2 = - M_n^2$ and arrive at the equation of motion
\begin{equation} \partial_r \left[r^{1+2s}\partial_r V\right] +{r^{2s-3} } M_n^2 V = 0 \label{eom} \end{equation}

The large $r$ solutions  take the form
\begin{equation} V = c + {c'\over r^{2s}} \end{equation}
The constant $c'$ has dimension $2+s$ which is appropriate to describe the operators under discussion and $c$ has dimension $2-s$ as is appropriate for the source. 

To provide intuition as to the form of the solutions we move to a Schroedinger equation form by setting $z= 1/r$ and $V=z^{s-1/2} \psi$ giving
\begin{equation}
- \psi^{''} + U(z) \psi =  M_n^2 \psi,  \hspace{1cm} U(z) =  {(s^2-1/4) \over z^2}\end{equation}
At small $z \rightarrow 0$ (the UV) the potential grows sharply and in the IR the hard wall presents another barrier - this is asymptotically a square well and the eigenstates $M_n^2 \sim n^2$. Further since the well is proportional in magnitude to $s^2$ it also follows that $M_s^2 \sim s^2$.

In soft wall models the IR metric and/or dilaton are allowed to take different forms to replace the hard wall cut off.  If we consider an IR metric ($z \rightarrow \infty$) 
\begin{equation} 
ds^2 = e^{2A(z)} (dx_4^2 + dz^2) \end{equation}
then $e^A$ now carries energy dimension and the action for our fields are
\begin{equation} S \sim \int d^4x dr ~~ \sqrt{-g} e^{-\Phi} {e^{(4s-4)A} \over 4 g_5^2} (\partial^\mu A^{\nu \lambda...})^2 \label{thin} \end{equation}
The equation of motion becomes 
\begin{equation} \partial_z \left[e^{(2s-1)A- \Phi}\partial_z V\right] +{e^{(2s-1)A - \Phi} } M_n^2 V = 0  \label{eom2} \end{equation}
The choices $\Phi=$ constant, $A=-z^2$ and the redefinition of fields $V = e^{(s-1/2)z^2} \psi$ gives the Schroedinger form
\begin{equation}
- \psi^{''} + U(z) \psi =  M_n^2 \psi,  \hspace{1cm} U(z) =  (4 s^2+1) z^2\end{equation}
Here the potential is that of a simple harmonic oscillator and the solutions are known to scale as $M_n^2 \sim n$. However, the $s$ dependence in the potential means that the behaviour with $s$ is not $M_s^2 \sim s$.

Alternatively one can return to (\ref{thin}) leaving the metric untouched and generate the soft wall through a dilaton profile $\Phi \sim 1/r^2 \sim z^2$. The equation of motion is that in (\ref{eom2}) with $A=-\log z$.
On setting $V= e^{(z^2-(2s-1)A)/2} \psi$ we find
\begin{equation}  
- \psi^{''} + U(z) \psi =  M_n^2 \psi,  \hspace{0.4cm} U(z) =   z^2 + 2(s-1) + {s^2 -1/4 \over z^2} \end{equation}
which has eigenvalues
\begin{equation}
m_{n,s}^2 = 4 (n+s) \end{equation}
which is the preferred scenario in \cite{Karch:2006pv}. 

Our initial goal is now to realize these soft wall scenarios in the Dynamic AdS/QCD model.

\section{Dynamic AdS/QCD}

Dynamic AdS/QCD was introduced in detail in \cite{Alho:2013dka} and is a variant of AdS/QCD with some additional features taken from top down D7 probe models. In particular the soft wall in the model is dynamically determined and corresponds to the presence of a quark condensate. 

The five dimensional action of our effective holographic theory is
\bea
S & = & \int d^4x~ d \rho\, e^{-\Phi} {\rm{Tr}}\, \rho^3 
\left[  {1 \over \rho^2 + |X|^2} |D X|^2 \right. \nonumber \\ 
&& \left.+  {\Delta m^2 \over \rho^2} |X|^2   + {1 \over 2} F_V^2  \right], 
\label{daq}
\eea
The  field $X$ describes
the quark condensate degree of freedom. Fluctuations in $|X|$ around its vacuum configurations describe the scalar meson. The $\pi$ fields are the phase of $X$,
\begin{equation} X = L(\rho)  ~ e^{2 i \pi^a T^a} .
\end{equation}
$F_V$ are vector fields that will describe the vector ($V$)  mesons. 

We work with the five dimensional metric 
\begin{equation} 
ds^2 =  { d \rho^2 \over (\rho^2 + |X|^2)} +  (\rho^2 + |X|^2) dx^2, 
\end{equation}
which will be used for contractions of the space-time indices.
$\rho$ is the holographic coordinate ($\rho=0$ is the IR, $\rho \rightarrow \infty$ the UV)
and $|X|=L$ enters into the effective radial coordinate in the space, i.e. there is an effective $r^2 = \rho^2 + |X|^2$. This is how the quark condensate generates a soft IR wall for the linearized fluctuations that describe the mesonic states: when $|X|$ is nonzero the theory will exclude the deep IR at $r=0$. 

The vacuum structure of the theory can be determined by setting all fields except $|X|=L$ to zero. We assume that $L$ will have no dependence on the $x$ coordinates. The action for $L$  is given by
\begin{equation} \label{act} S  =  \int d^4x~ d \rho ~  \rho^3 \left[   (\partial_\rho  L)^2 +  \Delta m^2 {L^2  \over \rho^2 }   \right].
\end{equation}
If $\Delta m^2 =0$ then the scalar, $L$, describes a dimension 3 operator and dimension 1 source as is required for it to represent $\bar{q} q$ and the quark mass $m$. That is, in the UV the solution for the $L$ equation of motion is $L = m + \bar{q}q/\rho^2$. This case is in fact the ${\cal N} =2$ SYM theory of \cite{Kruczenski} and Dynamic AdS/QCD generates the known spectrum for the $\rho$ meson in that case.  A non-zero $\Delta m^2$ allows us to introduce an anomalous dimension for the quark bilinear operator. If the mass squared of the scalar violates the BF bound of -4 ($\Delta m^2=-1$, $\gamma=1$) then the scalar field $L$ becomes unstable and the theory enters a chiral symmetry breaking phase. A controlled example is introducing a magnetic field into the ${\cal N} =2$ SYM theory \cite{Filev:2007gb} by an effective dilaton factor
\begin{equation} e^{-\Phi} = \sqrt{1 + {B^2 \over (\rho^2 + L^2)^2}} \end{equation}
One could alternatively expand this to quadratic order in $L$ and treat the quadratic term as a $\rho$ dependent $\Delta m^2$ term. The induced $L(\rho)$ function is schematically of the form $L \sim 1/(1 + \rho^2)$ for $B \sim 1$. This function can be thought of as the RG flow of the quark mass from a current quark mass of zero in the UV to a non-zero constituent quark IR value. 

Our immediate goal in this paper is not to fix $\Delta m^2$ and derive $L(\rho)$ but to investigate the form of $L(\rho)$ and $\Phi(\rho)$ which will give soft wall behaviour for the $\rho$ and its Regge tower of bound states.

\subsection{The $\rho$ Regge Trajectory}

We again assume the $s=1$ $\rho$ meson is created by the operator $\bar{q} \gamma^\mu q$, with higher Regge states associated with the operators $\bar{q} \gamma^\mu \partial^\nu q$,
$\bar{q} \gamma^\mu \partial^\nu \partial^\lambda q$,.. and so on with $s-1$ derivatives inserted. 

Holographically each of these operators will be associated with an $s$ index field in the geometry with action
\begin{equation} S \sim \int d^4x ~ d\rho ~~ {\rho^{1 + 2s} e^{-\Phi} \over (\rho^2 + L^2)^{1-s}}(\partial^\mu A^{\nu \lambda...})^2\end{equation}
Note there is some ambiguity here about where factors of $\rho$ or factors of $\sqrt{\rho^2 + L^2}$ occur (they are fixed by top down models \cite{Kruczenski} only for the $\rho$ meson) but for the generic conclusions we reach below this will not be important. 

We seek linearized solutions of the form $V(r) \epsilon^{\nu \lambda...} e^{-ik.x}$ with $k^2 = - M_n^2$ and arrive at the equation of motion
\begin{equation} \partial_\rho \left[ \rho^{1+2s}e^{-\Phi} \partial_\rho V\right] +{\rho^{1+2s} e^{-\Phi} \over (\rho^2 + L^2)^2} M_n^2 V = 0 \end{equation}

The large $\rho$ solutions (assuming $e^{-\Phi}$ becomes a constant) again take the form
$V = c + {c'\over \rho^{2s}}$.
We now seek to place the equation into Schroedinger form so that we can easily understand the potential that generates the masses of the tower of radially excited states. We first change coordinates so the $\partial_\rho^2 V$ term and the $M_n^2 V$ terms have the same coefficient
\begin{equation} \label{chco} \partial_\rho = {1 \over \rho^2 + L^2} \partial_z \end{equation}
giving
\begin{equation} \partial_z \left[ {\rho^{1+2s}e^{-\Phi} \over \rho^2 + L^2} \partial_z V\right] +{\rho^{1+2s} e^{-\Phi} \over \rho^2 + L^2} M_n^2 V = 0 \end{equation}

Next we rescale $V$
\begin{equation} \psi = a V \end{equation}
Requiring that the linear derivative term vanishes leads to
\begin{equation}
{1 \over a} \partial_\rho a = -{1+2s \over 2 \rho} + {1 \over 2} \partial_\rho \Phi + {\rho \over \rho^2 + L^2} + {(\partial_\rho L^2) \over 2 ( \rho^2 + L^2)} \end{equation}  
and hence
\begin{equation} \begin{array}{ccl}
{1 \over a} \partial_\rho^2 a &=& {1+2s \over 2 \rho^2} + {1 \over 2} \partial_\rho^2 \Phi + {1 \over \rho^2 + L^2} + {(\partial_\rho^2 L^2) \over 2 ( \rho^2 + L^2)} \\
&&\\&&- {(\partial_\rho L^2) (2 \rho + (\partial_\rho L^2)) \over 2 (\rho^2 + L^2)^2}- {\rho (2 \rho + (\partial_\rho L^2))\over (\rho^2 + L^2)^2}  + \left({1 \over a} 
\partial_\rho a \right)^2 \end{array}\end{equation}  

The equation of motion is now of Schroedinger form
\begin{equation} - \psi^{''} + U \psi = M_n^2 \psi \end{equation}
with the potential
\begin{equation} \begin{array}{ccl} 
U & = & - {(\rho^2+L^2)^2 \over a \rho^{1+2s} e^{-\Phi}} \partial_\rho \left[\rho^{1 + 2s} e^{-\Phi} \partial_\rho a \right] \\&&\\
&=& -(\rho^2 + L^2)^2 \left[ {(1+2s) \over \rho} \left({1 \over a} \partial_\rho a\right) - \partial_\rho\Phi \left({1 \over a} \partial_\rho a\right) + {1 \over a} \partial_\rho^2 a\right]\end{array}\end{equation}
This expression is in the $\rho$ coordinates and needs to be rewritten in $z$ using the result of the coordinate transformation in (\ref{chco})

\subsection{Examples}

In this section we will look at the spectrum of the $\rho$ meson and its Regge partners which emerge from a variety of choices of the function $L(\rho)$ in Dynamic AdS/QCD. We will begin in the controlled case of the ${\cal N}=2$ SYM theory and move away phenomenologically towards more QCD-like spectra. 

\subsubsection{${\cal N}=2$ SQCD}

The model includes the D3/probe D7 model which describes an ${\cal N}=2$ gauge theory \cite{Kruczenski}. In this case $L$ is a constant (which, up to a constant, is the quark mass and the only scale of the theory). The change of variables in (\ref{chco}) gives for $m=1$
\begin{equation}  \partial_\rho = {1 \over \rho^2 + 1} \partial_z, \hspace{1cm} z = \arctan[ \rho] \end{equation}
The key point here is that for $0 < \rho < \infty$ maps to $0 < z < \pi/2$. Since $z$ is of restricted range a square well like potential asymptotically is unavoidable. This is directly related to the fact that in a D7 probe model the $\rho$ meson physics lives on the D7 world-volume and does not access scales below the quark mass (here $r<1$). In the field theory the constituent quark mass provides a cut-off in RG scale and the meson masses are determined only in the theory above that cut-off.

In Fig. 1 we show the Schroedinger wells generated for different $s$ - they give the known analytic spectrum $M^2 = 4 (n + s) (n + s + 1)$. Some sample masses are also plotted showing $M^2$ grows as $n^2, ~ s^2$.

\begin{center}
\includegraphics[width=6.5cm]{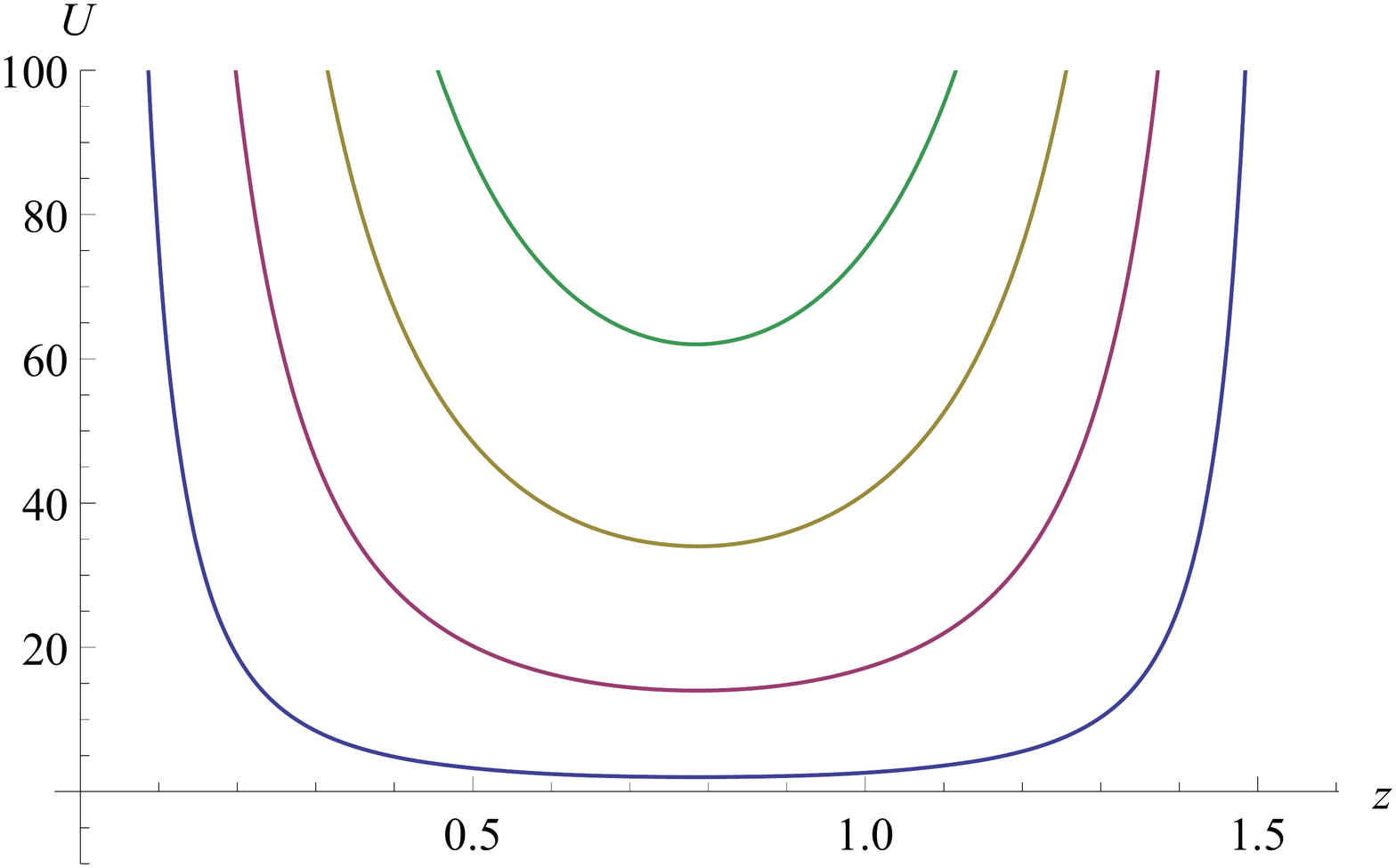}

\includegraphics[width=6.5cm]{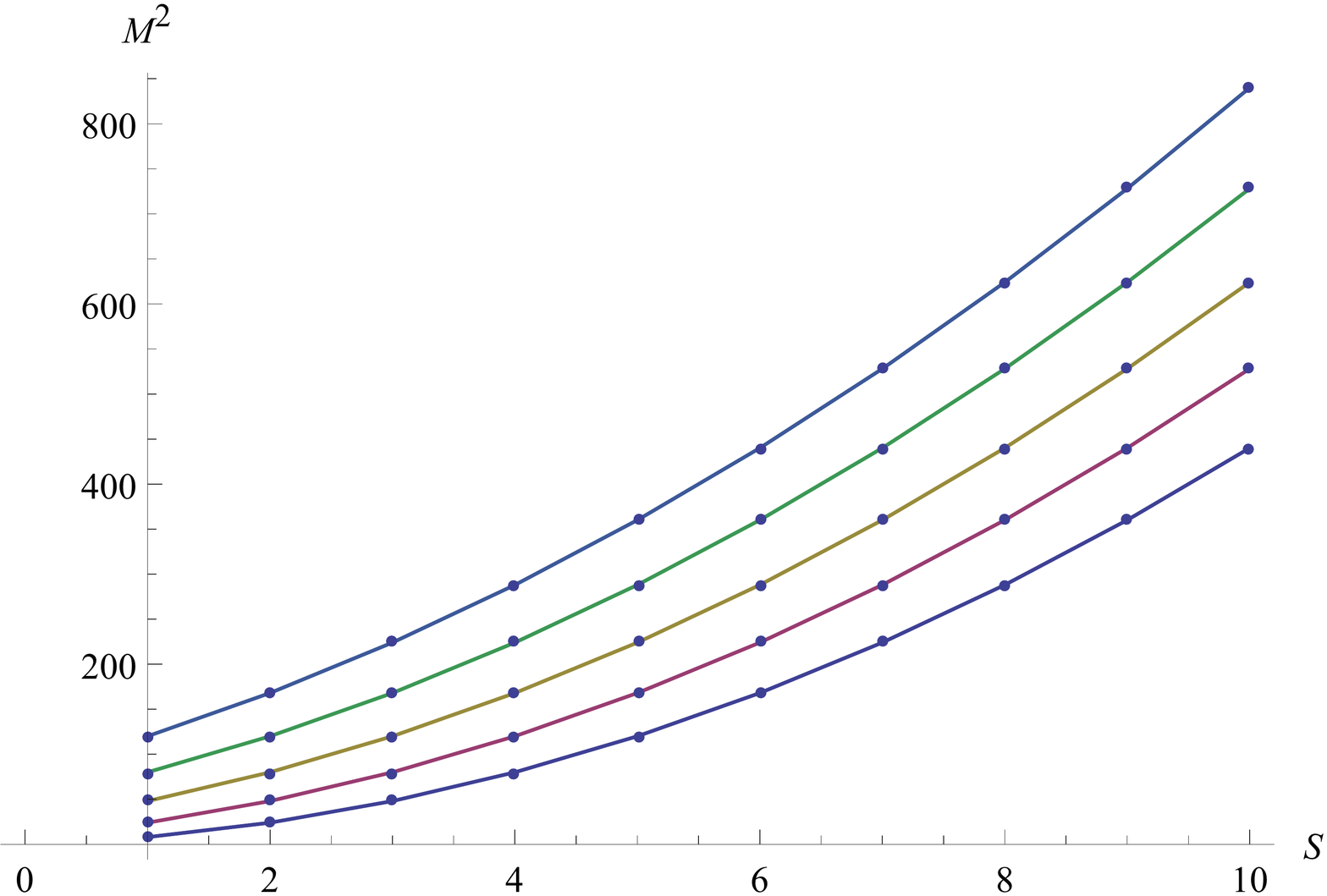}

Figure 1: The Schroedinger wells for the ${\cal N}$=2 model for $s=1,2,3,4$ which give a spectrum $M^2 = 4 (n + s) (n + s + 1)$,
and a plot of the Mass trajectories vs spin, $s$, for  the excitation numbers $n=1,2,3,4,5$.

\end{center}

\subsubsection{A Model of a Dynamically Generated Mass}

In models with a dynamically generated mass (eg \cite{Filev:2007gb,Babington:2003vm}) a typical profile for the embedding $L$ is
\begin{equation} 
L = {1 \over 1 + \rho^2}
\end{equation}

which falls off as $1/\rho^2$ at large $\rho$ but deviates from passing through $r^2=\rho^2+ L^2=0$  yet has $\partial_\rho L(0)=0$. Such models are very similar to the ${\cal N}=2$ case though in that at large $\rho$ the $L$ dependence in (\ref{chco}) is negligible whilst at small $\rho$, $L\simeq$ constant. Again one finds that the $z$ coordinate is bounded in extent and the Schroedinger well must be square asymptotically. In Fig. 2 we again plot the Schroedinger well and masses for this case. $M^2_{n,s}$ again grows as $n^2$ and $s^2$.  

Of course this is a toy example because the running of dilaton like factors that might induce this shape in $L$ are not included. However it is important to stress that the change of variables to $z$ is independent of the dilaton - it is this change of variables that leads to a truncated range in $z$ and hence a square well and $n^2$ like spectrum. No choice of dilaton in the bulk or on the brane could change that result.

\begin{center}
\includegraphics[width=6.5cm]{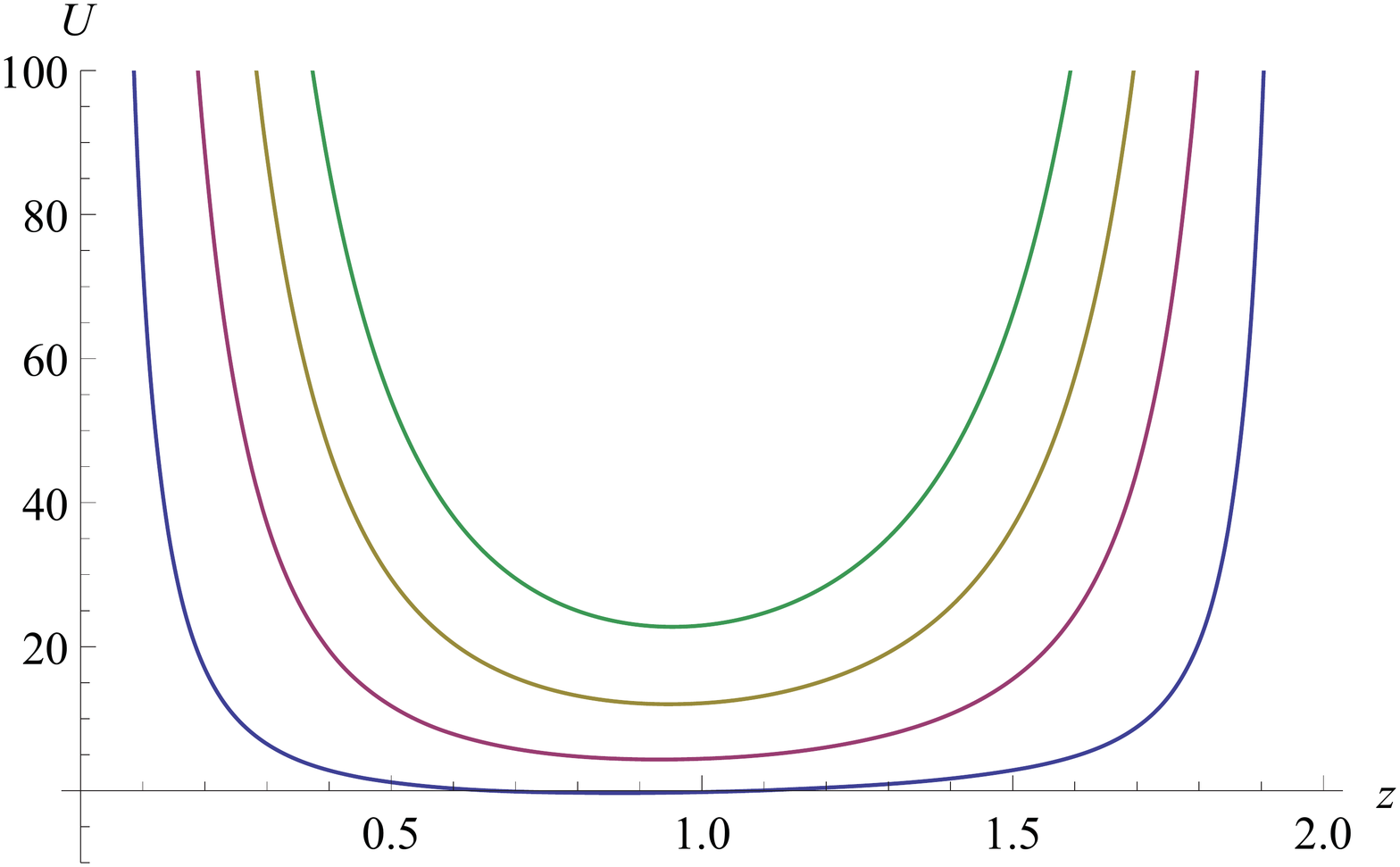}

\includegraphics[width=6.5cm]{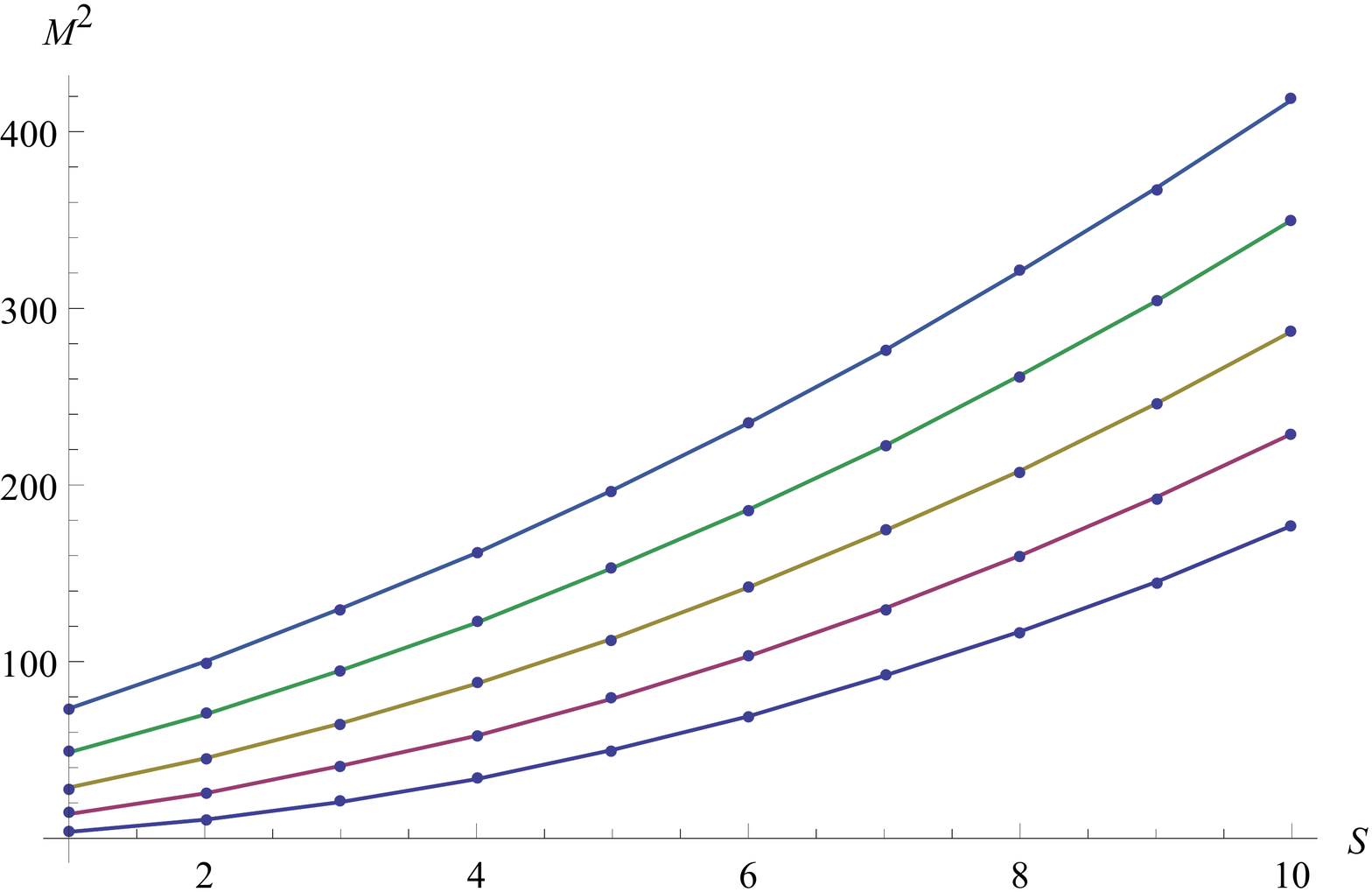}

Figure 2: The Schroedinger wells for the Dynamically Generated Mass model for $s=1,2,3,4$ 
and a plot of the Mass trajectories vs spin, $s$, for  the excitation numbers $n=1,2,3,4,5$.

\end{center}

\subsubsection{Engineering with $L$}

Now we can try to engineer a behaviour for the meson masses that goes as $\sqrt{n}$ or $\sqrt{s}$. Let's first ask what choice of $L(\rho)$, which sets the form of the soft wall,  would achieve this. We need as $\rho \rightarrow 0$ for $L$ to dominate in the factor $(\rho^2 + L^2)$ in (\ref{eom})  - if $L \sim \rho^p$ then we need $0 <p < 1$. For $p< 1/2$ the $z$ coordinate resulting from (\ref{chco}) is bounded by a maximum value and the asymptotics must look like a square well. For $p>1/2$ the IR potential well falls to zero and the spectrum is not discrete. More interesting behaviours can be found in the region fine tuned close to $p=1/2$. 

We can for example engineer the softwall potential of \cite{Karch:2006pv} for the $\rho$ mesons. We set $\Phi$ = constant and $s=1$. We enforce the change of variables
\begin{equation} e^{-z^2} \partial_z = -\rho^3 \partial_\rho \end{equation}
which at large $z$, small $\rho$ implies
\begin{equation} \rho^2 = z e^{-z^2} \end{equation}
We get the equation of motion in the IR limit where $L$ dominates its term

\begin{center}
\includegraphics[width=6.5cm]{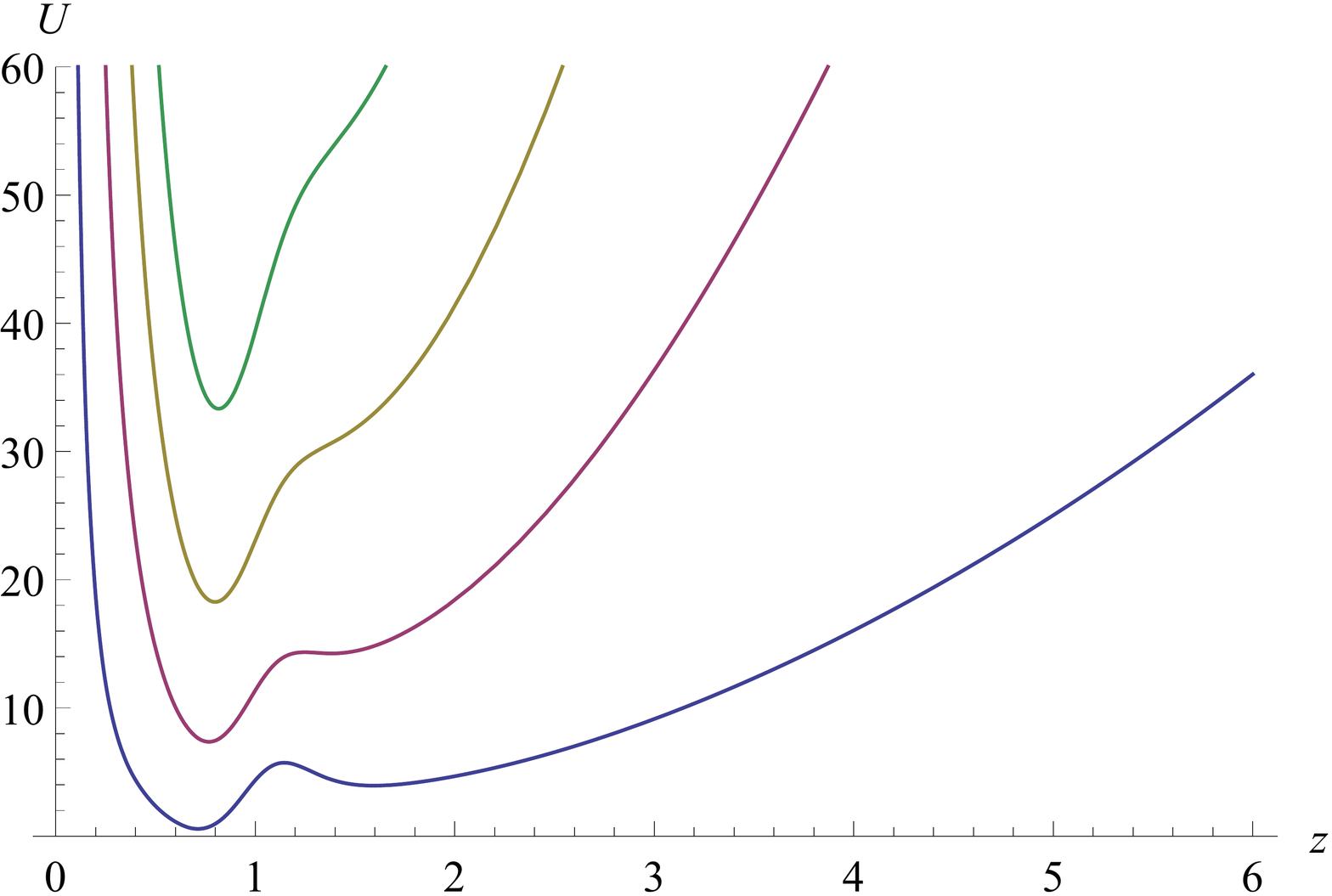}

\includegraphics[width=6.5cm]{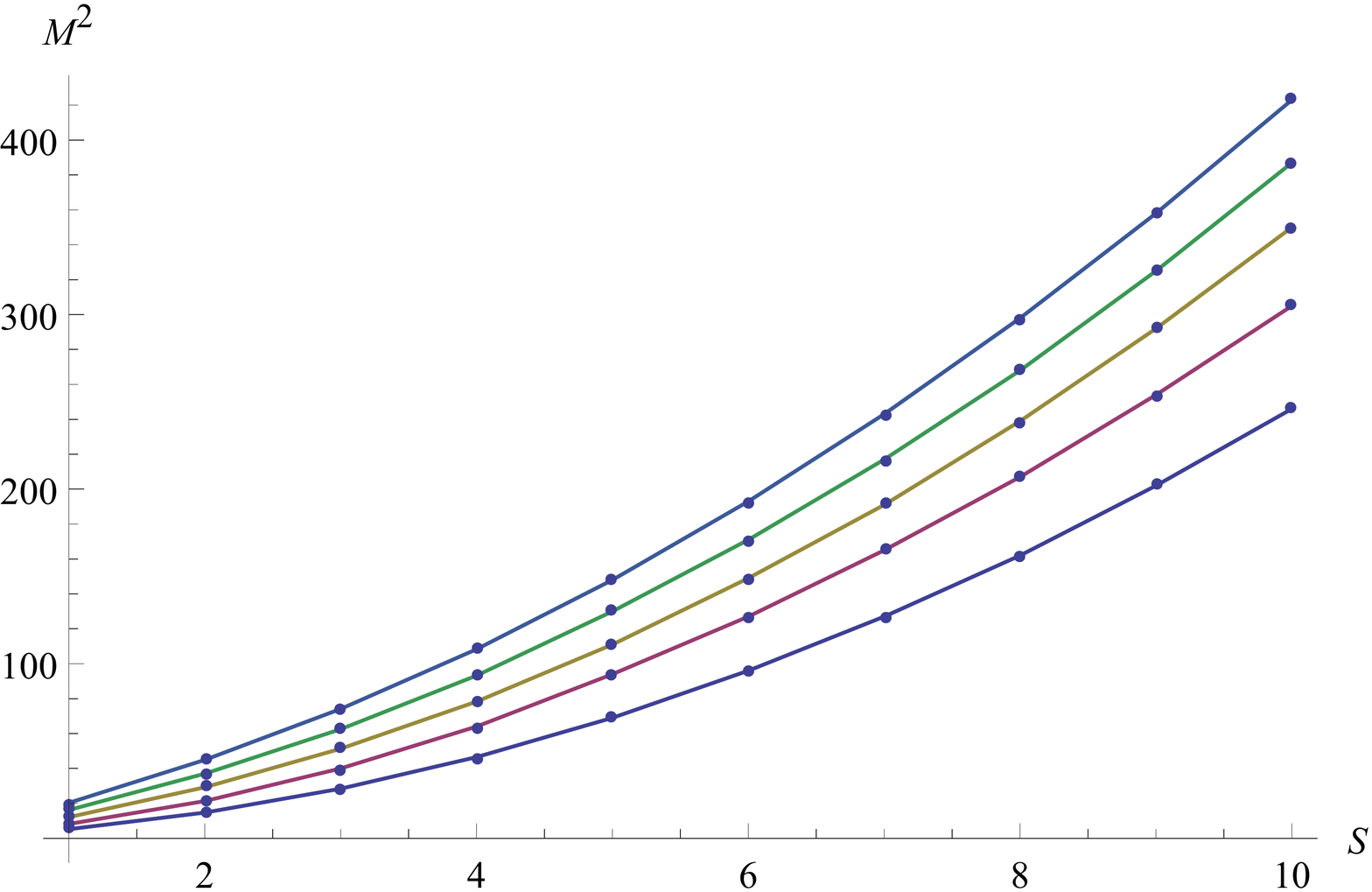}

Figure 3: The Schroedinger wells for the softwall model with $L$ given by (\ref{guess}) for $s=1,2,3,4$ 
and a plot of the Mass trajectories vs spin, $s$, for  the excitation numbers $n=1,2,3,4,5$.
\end{center}

\begin{equation}
\partial_z ( e^{-z^2} \partial_z V) + {\rho^6 e^{z^2} \over L^4} V = 0 
\end{equation}
Thus if we pick
\begin{equation} {\rho^6 e^{z^2} \over L^4}  = e^{-z^2}, \hspace{1cm} {ie}~~~~~ L^2 = z \rho \end{equation}
in the IR, we achieve the softwall model of (\ref{eom2}) at $s=1$. Note up to a log factor we indeed sit on the $p=1/2$ boundary. 

As an example complete model of this type we can choose
\begin{equation}
L = { z^{1/2} \rho^{1/2} \over \sqrt{1 + z \rho^5}   } \label{guess}    \end{equation}
This falls off asymptotically as $\rho^2$ but matches the IR behaviour needed.

We display the Schroedinger well and Regge trajectories in Fig. 3. 
The potential has a harmonic oscillator form at large $z$ which leads to the linear Regge behaviour in $n$. 
Is this a success? First let's plot the function $L(\rho)$  - see Fig. 4.

\begin{center}
\includegraphics[width=6.5cm]{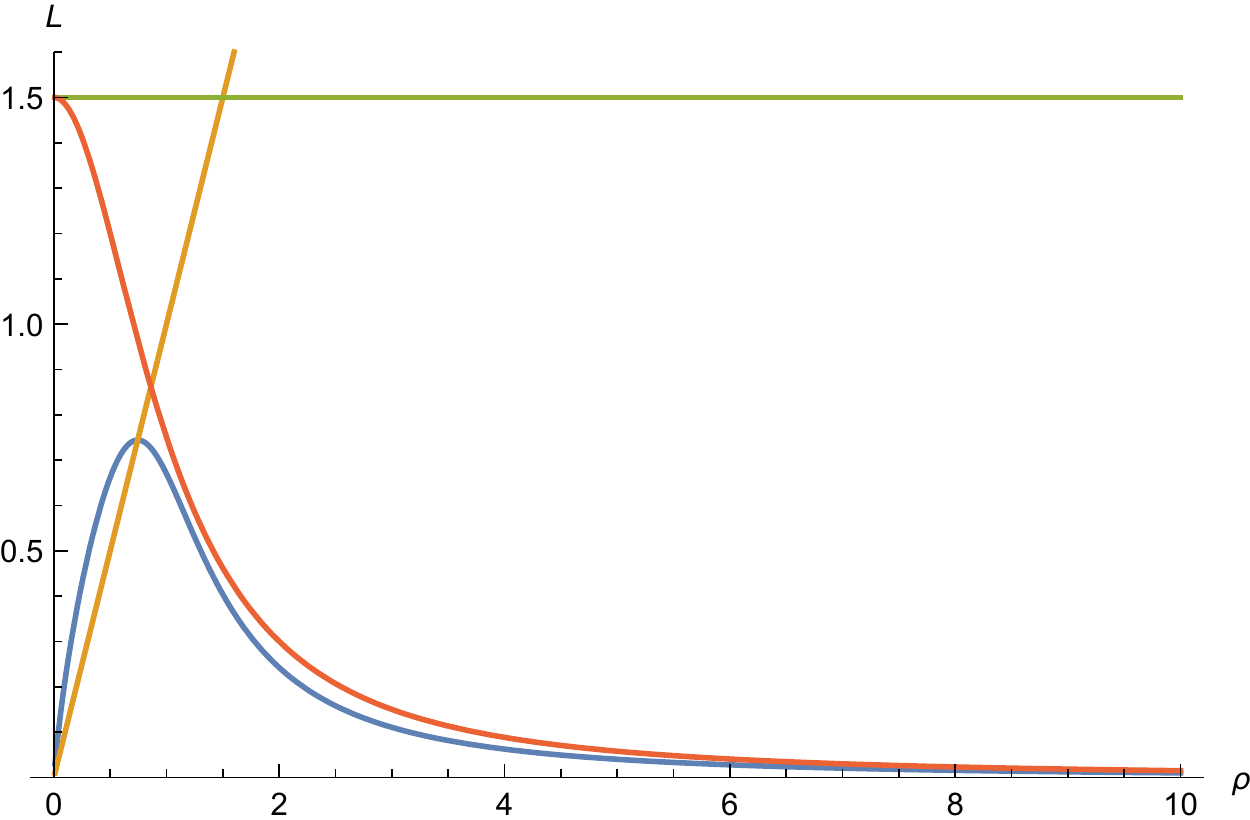}

Figure 4: The function $L(\rho)$ which reproduces the softwall behaviour of \cite{Karch:2006pv} with a constant dilaton. Also shown are examples of the profiles for the ${\cal N}=2$ theory ($L=$constant) and for the dynamically generated mass example ($L=1/(1+\rho^2)$). The line $L=\rho$ is also plotted to show where the on-mass shell condition is satisfied.
\end{center}

Physically $L(\rho)$ in the top down models is a plot of the quark mass against RG scale. Here this function is very peculiar, at least in the context of top down models. The quark mass grows until the on-mass shell scale but then below that scale falls to zero in the deep IR. In particular in this construct the $\rho$ meson physics is determined by radial distances (RG scales) all the way down to zero. This is in sharp contrast to top down models where the $\rho$ physics is immune to scales below the IR quark mass. The construct of a soft wall needs non-decoupling of quarks in the IR of a strongly coupled gauge theory. Of course this generic point is true in any softwall model. Many people have expressed the view that softwalls are not the way to produce linear Regge behaviour (one should use true stringy behaviour) and the interpretation in Dynamical AdS/QCD probably supports this view.  

Functionally there is a second problem with this model. By manipulating $L$ we have effectively realized linear trajectories in an equivalent way to the use of ``$e^{A}$'' in \cite{Karch:2006pv}. As there, the trajectories for higher $s$ states do not have the same slope as $s=1$. In the next example we will provide a model that mixes a dilaton flow and $L$ profile that achieves the best case of \cite{Karch:2006pv}.

\begin{center}
\includegraphics[width=6.5cm]{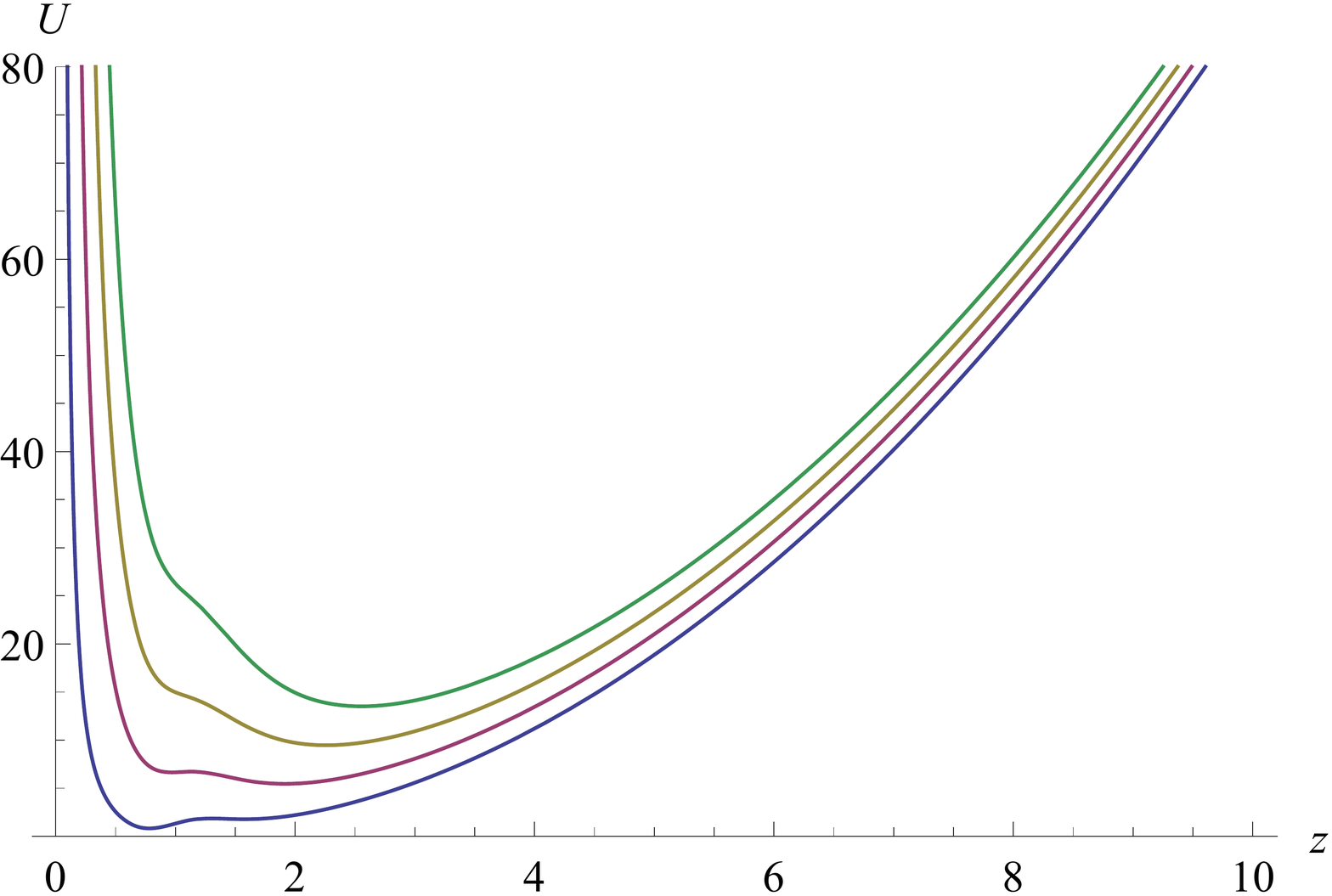}

\includegraphics[width=6.5cm]{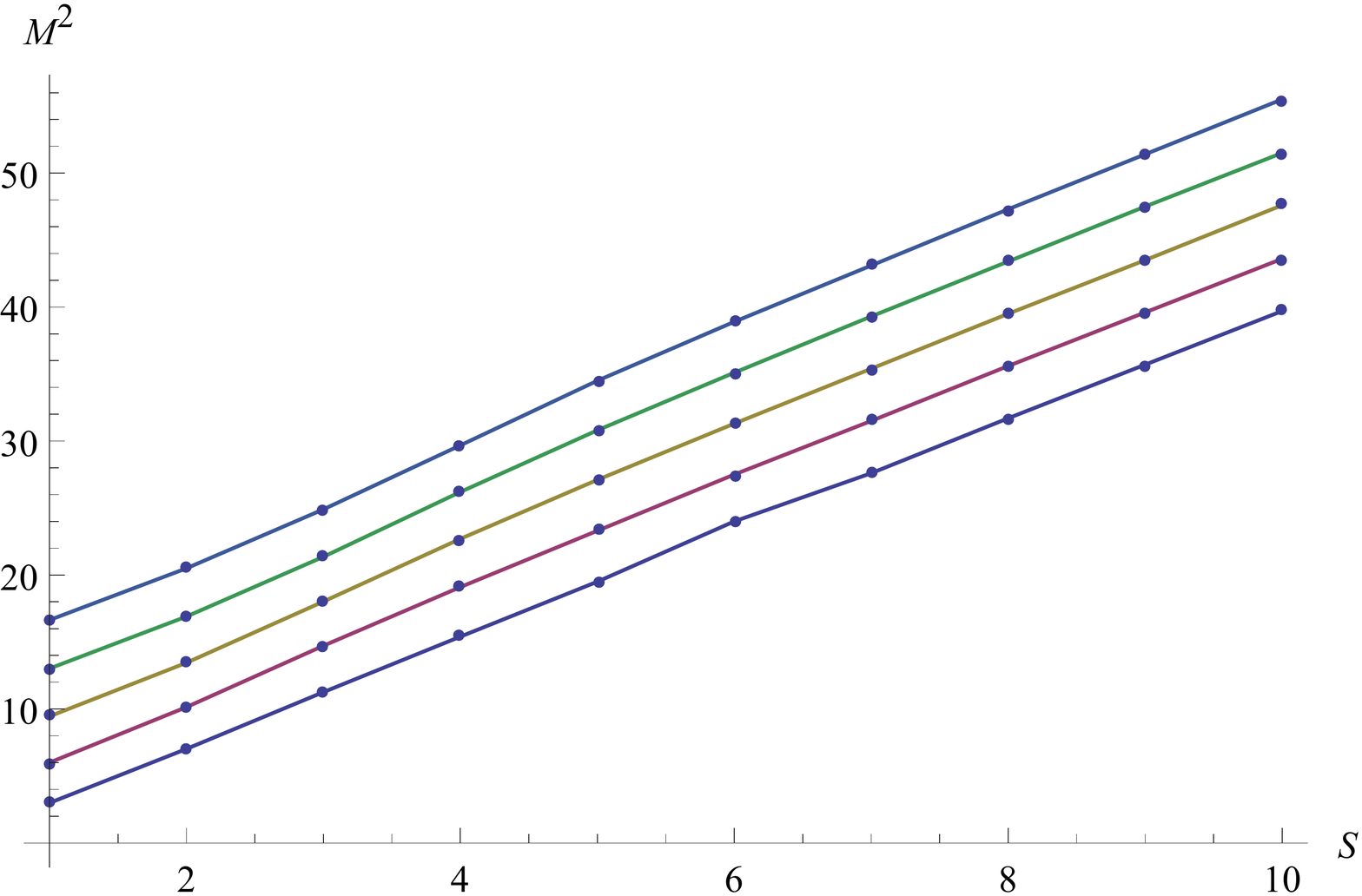}

Figure 5: The Schroedinger wells for the softwall model with $L$ given by (\ref{guess}) for $s=1,2,3,4$ 
and a plot of the Mass trajectories vs spin, $s$, for  the excitation numbers $n=1,2,3,4,5$.
\end{center}

\subsubsection{Engineered Dilaton and $L$}

As we have seen in our second example we can't use a bulk dilaton to engineer an IR softwall model with any finite value for $L(0)$ because the quark physics will not see the deep IR behaviour of that dilaton. An example solution to this problem is to take $L \sim \rho$ in the IR so that the induced metric on the embedding is just AdS and take a dilaton $\phi = z^2$. We set as an example
\begin{equation} \label{guess} L = {\rho \over 1 + \rho^3} , \hspace{1cm} \Phi = z^2 \end{equation}      
We plot the Schroedinger wells and mass trajectories in Fig. 5 where the linearity in $n$ and $s$ is clear. 

The complaint that the $\rho$ physics is determined by RG scales below the IR quark mass remains in this model.

Note that we have not attempted to find profiles for $\Delta m^2$ which would dynamically generate these profiles for $L$. They would clearly need to be quite peculiar relative to the standard expectation from the perturbation theory running.

\section{Techni-Dilaton}

In this final section we want to turn our attention to the impact of soft wall dynamics on the so called techni-dilaton state in walking gauge theories. There is considerable interest in the behaviour of QCD-like SU($N_c$) gauge theories with varying number of flavours $N_f$ \cite{Caswell:1974gg,Banks:1981nn,Appelquist:1996dq,Appelquist:1998rb,Appelquist:1988yc,Ryttov:2007cx,Ryttov:2007sr,Dietrich:2006cm,Sannino:1999qe,Armoni:2009jn,Gies:2005as}. For $N_f < 11 N_c/2$ the theories become asymptotically free. There is believed to be a region of $N_f$ below this value where the theory runs to an IR fixed point at which the coupling grows in strength as $N_f$ decreases. At some critical value, roughly estimated as $N_f \simeq 4 N_c$, the coupling at the fixed point becomes strong enough to trigger chiral symmetry breaking and IR conformality is lost. The transition is triggered when the anomalous dimension of $\bar{q} q$ $\gamma \simeq 1$ \cite{Jarvinen:2011qe, Kutasov:2012uq, Alvares:2012kr}. The transition is of a BKT or Miransky scaling type \cite{Miransky:1996pd,Kaplan:2009kr}. Just below the critical value of $N_f$ for chiral symmetry breaking theories are supposed to display walking behaviour. The coupling runs to an IR theory with $\gamma \simeq 1$ from below. For theories close to the critical coupling the running in the IR theory is very slow and the scale where chiral symmetry breaking occurs is in a theory that is very close to conformal \cite{Holdom:1981rm}.  A number of authors have predicted that a light dilaton-like state (relative to the rest of the spectrum) will emerge in these theories \cite{Sundrum:1991rf,Appelquist:1998xf,Yamawaki:1985zg,Bando:1986bg,Hong:2004td,Dietrich:2005jn}.

Recently several groups have developed holographic models of walking theories and the conformal window \cite{Jarvinen:2011qe, Kutasov:2012uq, Alho:2013dka}. These theories do not predict the dynamics of the running of the couplings or anomalous dimensions but include them either directly or through chosen potentials for supergravity fields. They do though predict the meson spectrum as a function of $N_f$ and $N_c$ after those assumptions have been included.  Interestingly Dynamic AdS/QCD was used to show the presence of a light $\bar{q} q$ scalar state \cite{Alho:2013dka} but the model of \cite{Jarvinen:2011qe} does not see such a state. Here we want to argue that it is the incorporation of a soft wall dynamic in \cite{Jarvinen:2011qe} that explains this difference. In particular, as we have seen in the sections above, to achieve $M_n^2 \sim n$ Regge trajectories in AdS/QCD models it is necessary to allow the meson physics to be determined by the deep IR regime below the chiral symmetry breaking scale. In \cite{Jarvinen:2011qe} this is achieved by the tachyon field that describes the $\bar{q}q$ condensate diverging in the IR only at the scale $r=0$. In Dynamic AdS/QCD the probe-brane-like  action terminates at the on mass shell condition for the quarks. The running in the gauge theory is near conformal down to that on mass-shell condition scale but below where the quarks decouple from the running of the gauge coupling the Yang-Mills like running of the glue theory is very non-conformal. We believe that whether a light mesonic state is seen or not depends on whether its dynamics is sensitive to the conformal symmetry breaking deep IR or not. To demonstrate this logic we are simply going to study the mass of the scalar $\bar{q}q$ state in Dynamic AdS/QCD with and without soft wall behaviour. When we decouple the mesonic action at the quark mass scale the state is light but if we allow it to see the non-conformal IR running then the state becomes heavy with mass of order the chiral symmetry breaking scale. This at least highlights the role of quark decoupling in these holographic descriptions. Top down probe brane models appear to us to support the idea that mesonic physics should be blind to the deep IR below the quark mass but this would invalidate the softwall mechanism.

\subsection{Walking Dynamics in Dynamic AdS/QCD}

The gauge dynamics is input into Dynamic AdS/QCD through the running of the anomalous dimension of $\bar{q} q$, $\gamma$. In the holographic description $\gamma$ enters through the term $\Delta m^2$ in (\ref{daq}).

We will fix the form of $\Delta m^2$ using the two loop running of the gauge coupling in QCD with $N_f$ flavours transforming under a representation $R$. This of course is a naive extrapolation of perturbative results beyond their regime of validity but is widely used to motivate the presence of a conformal window and walking, and because we have no better guess.  The running takes the form
\begin{equation} 
\mu { d \alpha \over d \mu} = - b_0 \alpha^2 - b_1 \alpha^3,
\end{equation}
where
\begin{equation} b_0 = {1 \over 6 \pi} (11 N_c - 2N_f), \end{equation}
and
\begin{equation} b_1 = {1 \over 24 \pi^2} \left(34 N_c^2 - 10 N_c N_f - 3 {N_c^2 -1 \over N_c} N_f \right) .\end{equation}
Asymptotic freedom is present provided $N_f < 11N_c/2$. There is an IR fixed point with value
\begin{equation} \alpha_* = -b_0/b_1\,, \end{equation}
which rises to infinity at $N_f \sim 2.6 N_c$. 

The one loop result for the anomalous dimension of the quark mass is
\begin{equation} \gamma_1 = {3 C_2 \over 2\pi}\alpha, \hspace{1cm} C_2= {(N_c^2-1) \over 2 N_c} \,.  \end{equation}
So, using the fixed point value $\alpha_*$, the condition $\gamma=1$ occurs at $N_f^c \sim 4N_c$ (precisely $N_f^c = N_c \left( {100 N_c^2 - 66 \over 25 N_c^2 - 15}\right)$).

We will identify the RG scale $\mu$ with the AdS radial parameter $r = \sqrt{\rho^2+L^2}$ in our model. Note it is important that $L$ enters here. If it did not and the scalar mass was only a function of $\rho$ then, were the mass to violate the BF bound at some $\rho$, it would leave the theory unstable however large $L$ grew. Including $L$ means that the creation of a non-zero but finite $L$ can remove the BF bound violation leading to a stable solution. 

Working perturbatively from the AdS result $m^2 = \Delta(\Delta-4)$ we have
\begin{equation} \label{dmsq3} \Delta m^2 = - 2 \gamma_1 = -{3 (N_c^2-1) \over 2 N_c \pi} \alpha\, .\end{equation}
This will then fix the $r$ dependence of the scalar mass through $\Delta m^2$ as a function of $N_c$ and 
$N_f$.

The vacuum structure for a given choice of representation, $N_f$ and $N_c$ must be identified first. The Euler-Lagrange equation for the  vacuum embedding $L_v$ is given at fixed $\Delta m^2$ by the solution of 
\begin{equation}\label{embedeqn}
 \frac{\partial}{\partial\rho}\left( \rho^3 \partial_\rho L_v\right)  - \rho \Delta m^2 L_v =0.
\end{equation}
Note that if $\Delta m^2$ depends on $L_v$ at the level of the Lagrangian then there would be an additional term  $- \rho L^2 \partial \Delta m^2 / \partial L_v$. We neglect this term and instead impose the running of $\Delta m^2$ at the level of the equation of motion. The reason is that the extra term introduces an effective contribution to the running of $\gamma$ that depends on the gradient of the running coupling. Such a term is not present in perturbation theory in our QCD-like theories - we wish to keep the running of $\gamma$ in the holographic theory as close to the perturbative guidance from the gauge theory as possible.

In order to find $L_v(\rho)$ we solve the equation of motion numerically with shooting techniques with an input IR initial condition. A sensible first guess for the IR  boundary condition is
\begin{equation}\label{bca}  L_v(\rho=L_0) = L_0, \hspace{1cm}  L_v'(\rho=L_0)=0. \end{equation}
This IR condition is similar to that from top down models \cite{Babington:2003vm,Kruczenski:2003uq,Sakai:2004cn,Filev:2007gb} but imposed at the RG scale where the flow becomes ``on-mass-shell". Here we are treating $L(\rho)$ as a constituent quark mass at each scale $\rho$. To continue the flow below this quark mass scale we would need to address the  issue of the decoupling of the quarks from the running function $\gamma$ - previously we have not addressed this challenge but we will show some of the subtly below when we include a soft wall. 
\bigskip

The isoscalar $\bar{q}q$ ($\sigma$) mesons are described by linearized fluctuations of $L$ about its vacuum configuration, $L_v$. We look
for space-time dependent excitations,  ie $|X| = L_v + \delta(\rho) e^{i q.x}$,  $q^2=-M_\sigma^2$. The linearized equation of motion for $\delta$ is
\begin{equation} \label{deleom} \begin{array}{l}\partial_\rho( \rho^3 \delta' ) - \Delta m^2 \rho \delta -   \rho L_v \delta \left. \frac{\partial \Delta m^2}{\partial L} \right|_{L_V} \\  \\ \left. \right. \hspace{3cm}
+ M_\sigma^2 R^4 \frac{\rho^3}{(L_v^2 + \rho^2)^2} \delta  = 0 \end{array} \end{equation}
We seek solutions with, in the UV, asymptotics of $\delta=\rho^{-2}$ and with $\partial_\rho\delta|_{L_0}=0$ in the IR, giving a discrete meson spectrum. 

This calculation has already been presented in \cite{Alho:2013dka} and the $\sigma$ meson mass, in units of the $\rho$ meson mass, falls to zero as $N_f \rightarrow 4 N_c$ where the Miransky phase transition occurs. We summarize this result for the $N_c=3$ theory in Fig. 6.

\begin{center}
\includegraphics[width=6.5cm]{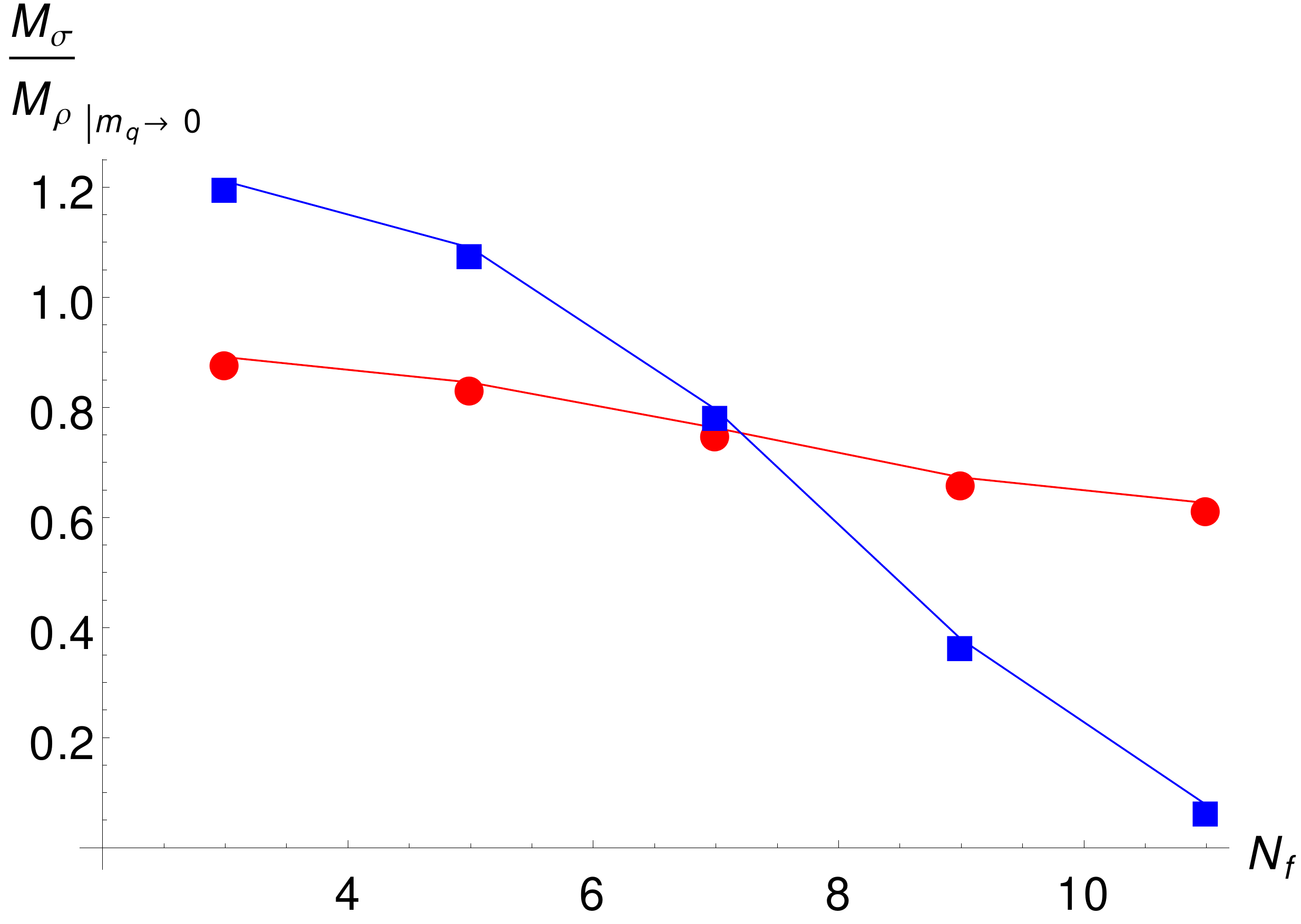}

Figure 6: A plot of the $\sigma$ meson mass in units of the $\rho$ meson mass against $N_f$ in Dynamic AdS/QCD. The points that fall to zero at the chiral transition at $N_f=12$ are those computed where the action is cut-off at the on-shell mass of the quark so the theory is blind to the deep IR conformal symmetry breaking, whilst the points that asymptote to a non-zero value at $N_f=12$ are for the soft wall variant.
\end{center}

Now we will include an IR soft wall behaviour into the model in the spirit of the fourth example given above with an IR $L(\rho)$ and  dilaton profile . To demonstrate the idea, rather than cooking $\Delta m^2$ below the quark on mass-shell scale, we will include the soft wall by hand in the $L_v$ profile and just compute the bound state masses. We take for the IR wall
\begin{equation} 
L(\rho) = L_0 \sin (\pi \rho  / 2 L_0), ~~~~~ \rho < L_0 \end{equation}
which is linear at small $\rho$ and matches to the solutions already found above $\rho=L_0$. We then include a dilaton 
of the form $1 + exp(-1/\rho^2) - exp(-1/L_0^2)$ which grows from unity at the matching scale to take the form $exp(-z^2)$in the IR. Now we again compute the $\rho$ and $\sigma$ spectrum as a function of $N_f$ for the $N_c=3$ theory except now imposing the boundary conditions $V'(0)=0$ and $\delta'(0)=0$. We display the results also in Fig 6. The $\sigma$ meson now does not become light because it is sensitive to the IR conformal symmetry breaking that picks out the scale $L_0$, as advertised.

\section{Summary}

Dynamic AdS/QCD is an AdS/QCD model derived from the quadratically expanded D3/probe-D7 system but with the running of the anomalous dimension of $\bar{q}q$ replaced by hand to match a particular theory. The field holographic to the quark bilinear $|X|=L$ can be thought of as the running quark mass and generates a soft wall in the action for the mesonic fluctuations. In this paper we have studied manipulating $L$ to mimic a soft wall model in the spirit of \cite{Karch:2006pv}. The idea is to use the deep IR behaviour of the metric or the dilaton to transform the mesonic spectra from that associated with a square well potential ($M_n \sim n$) to that associated with a simple harmonic oscillator ($M_n \sim \sqrt{n}$). Although this can be done, it requires  $L \rightarrow 0$ in the IR which is not the behaviour seen in top down models nor what would be naively expected for the IR behaviour of the quark mass. The mesonic physics becomes determined by RG scales well below the on-shell quark mass. Nevertheless we have shown examples of this behaviour in the model.  

We have also highlighted the consequences of soft wall dynamics in models of the light $\sigma$ meson that emerges in walking theories near the chiral transition from the conformal window. If the mesonic physics is determined by scales above the on-shell mass of the quark then the near-conformal dynamics of these theories can lead to a light $\sigma$ meson relative to the rest of the spectrum. However, if soft wall dynamics is included then the meson physics is determined at scales both above and below the on-shell mass. Since here the mass function falls to zero in the IR the dynamics is very non-conformal and the $\sigma$ meson mass is determined by the on-shell mass scale. This highlights the cause of the difference in spectrum between the results in \cite{Jarvinen:2011qe} and \cite{Alvares:2012kr}. Of course, in the holographic models the decoupling behaviour of the quarks or mesonic physics is input by hand so one must decide which scheme is most plausible. 

Our intuition is led by top down models such as the D3/probe D7 system with a magnetic field \cite{Filev:2007gb} where the mesonic physics lives on the probe and sees none of the geometry below the IR mass scale. Another example is the Sakai Sugimoto model \cite{Sakai:2004cn} in which, firstly, meson physics is restricted to the probe brane energy scales which are sharply cut off at the IR mass, and, secondly, the deep IR of the background geometry is capped off by confinement. We tend to feel the decoupling in \cite{Alvares:2012kr} is more realistic although one might expect some interaction with scales just below the quark mass that may tend to raise the techni-dilaton mass. Probably the lattice is the only way to find a definitive conclusion as to the fate of the techni-dilaton. 

\bigskip \bigskip \bigskip

\noindent{\bf Acknowledgements:} The authors are grateful for comments on the manuscript by Elias Kiritsis and Matti Jarvinen. We thank Alfonso Ballon-Bayona and Miguel Costa for discussions. NE thanks the Galileo Galilei Institute for Theoretical Physics for their hospitality and the INFN for partial support during the progress of this work. NE and MS are grateful for the support of STFC funding. PJ is supported by a University of Southampton STAG studentship.

\end{document}